\documentclass[aps,prl,showpacs,floatfix,twocolumn,superscriptaddress]{revtex4}
\usepackage{microtype} \usepackage{amssymb} \usepackage{amsmath}
\usepackage{graphicx}
\usepackage[latin1]{inputenc}
\begin{document}
\setlength{\parindent}{0pt} \newcommand{\llangle}{\langle \langle}
\newcommand{\rrangle}{\rangle \rangle}
%%%
\title{AC magnetization transport and power absorption in non-itinerant spin chains}
\author{Bj{\"o}rn Trauzettel}
\affiliation{Institute for Theoretical Physics and Astrophysics, University of
W{\"u}rzburg, D-97074 W{\"u}rzburg, Germany}
\affiliation{Department of Physics, University of Basel, CH-4056 Basel, Switzerland}
\author{Pascal Simon}
\affiliation{Department of Physics, University of Basel,
CH-4056 Basel, Switzerland}
\affiliation{Laboratoire de Physique et Mod{\'e}lisation des Milieux
  Condens{\'e}s, CNRS and Universit{\'e} Joseph Fourier, BP 166, F-38042
  Grenoble, France}
\author{Daniel Loss}
\affiliation{Department of Physics, University of Basel,
CH-4056 Basel, Switzerland}

\pacs{75.10.Jm, 75.40.Gb}

\begin{abstract}
We investigate the ac transport of magnetization in non-itinerant quantum systems such as spin chains described by the XXZ Hamiltonian. Using linear response theory, we calculate the ac magnetization current and the power absorption of such magnetic systems. Remarkably, the difference in the exchange interaction of the spin chain itself and the bulk magnets (i.e. the magnetization reservoirs), to which the spin chain is coupled, strongly influences the absorbed power of the system. This feature can be used in future spintronic devices to control power dissipation. Our analysis allows to make quantitative predictions about the power absorption and we show that magnetic systems are superior to their electronic counter parts.
\end{abstract}

\date{April 2008} \maketitle

Power dissipation is one of the most important limitations of state-of-the-art electronic systems. The same is true for spintronic devices in which spin transport is accompanied by charge transport. In non-itinerant quantum systems, the dissipation problem is reduced since true magnetization transport generates typically much less power than charge currents \cite{hall2006,ovchi2008}. This is one of the main reasons for putting so much hope and effort into spin-based devices for future applications \cite{wolf2001,awschalom2002,awschalom2007}.

\begin{figure}

\vspace{0.3cm}

\begin{center}
\leavevmode \includegraphics[width=7.5cm]{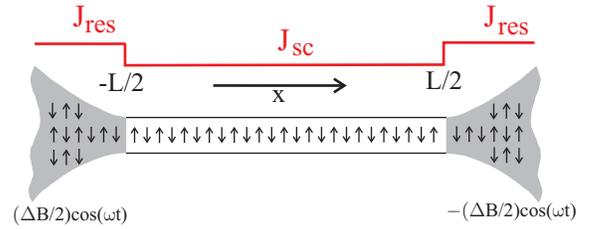}
\caption{(color online) Schematic of a quantum spin chain coupled to magnetization reservoirs. The magnetic field bias $\Delta B$ changes periodically in time. In the upper part of the figure, we illustrate that the exchange coupling in the spin chain $J_{\rm sc}$ (for $|x|<L/2$) can in general be different from the exchange coupling in the grey-shaded magnetization reservoirs $J_{\rm res}$. As suggested in Ref.~\cite{meier2003}, this setup can be realized by a bulk material with an intra-chain exchange much stronger than the inter-chain exchange, where the material is heated to a temperature $T > T_N$ in the central part and cooled to $T \ll T_N$ in the reservoir parts. ($T_N$ is the N\'eel ordering temperature.)}
\label{fig_setup}
\end{center}
\end{figure}

Here, we analyze non-itinerant quantum systems described by a spin Hamiltonian in which ac magnetization transport occurs via magnons or spinons (without the transport of charge). In Ref.~\cite{meier2003}, the spin conductance of such a device has been derived with a particular focus on the role of the magnetization reservoirs to which a one-dimensional spin chain is attached. We generalize this theory to the response to an ac magnetization source. This allows us to directly calculate (and thus estimate) the power absorption of such magnetic systems at a given driving frequency $\omega$ using linear response theory. In general, the exchange coupling $J$ in the spin chain and in the reservoirs will be different which is schematically illustrated in Fig.~\ref{fig_setup}. It turns out that the difference of the exchange coupling plays a crucial role in the dependence of the absorbed power as a function of $\omega$. The larger the difference the stronger will be the suppression of power dissipation at finite frequencies. At low frequencies, however, the dissipative power is independent of the difference of the exchange couplings and takes a universal value determined by $J$ in the reservoirs.

We analyze the ac transport problem in quantum spin chains by a mapping of the spin Hamiltonian coupled to magnetization reservoirs to the so-called inhomogeneous Luttinger liquid (LL) Hamiltonian \cite{maslov1995,ponomarenko1995,safi1995}. Interestingly, the absorbed power that is derived in this letter has an astonishingly simple dependence on the interaction parameters of the LL model, see Eq.~(\ref{W_result}) below. This makes it a prime candidate for the experimental observation of LL physics in nature.

%%%%%%%%%%%%%%%%%%%%%%%%%%%%%%%%%%%%%%%
%%%%%%%%%%%%%%%%%%%%%%%%%%%%%%%%%%%%%%%
%%%%%      FIGURE    SC Setup    %%%%%%
%%%%%%%%%%%%%%%%%%%%%%%%%%%%%%%%%%%%%%%
%%%%%%%%%%%%%%%%%%%%%%%%%%%%%%%%%%%%%%%

In order to describe the system shown in Fig.~\ref{fig_setup}, we consider a one-dimensional XXZ spin chain in the presence of a time-dependent magnetic field
$\mathbf{B} (\mathbf{x}_i,t) = B_i(t) \mathbf{e}_z$
which can be described by the Hamiltonian $H=H_{\rm XXZ}+H_B(t)$, where
\begin{eqnarray}
H_{\rm XXZ} &=&  J \sum_{\langle i,j \rangle} \Bigl( s_{i,x} s_{j,x} + s_{i,y} s_{j,y} + \Delta
s_{i,z} s_{j,z} \Bigr) , \\
H_B(t) &=& g_e \mu_B \sum_i B_i(t) s_{i,z} .
\end{eqnarray}
Here, $s_{i,\alpha}$ is the $\alpha$-component of the spin operator at
$\mathbf{x}_i$, $\langle i,j \rangle$ denotes nearest neighbor sites, $g_e$ is the g-factor, $\mu_B$ Bohr's magneton, and we assume
anti-ferromagnetic coupling with $J,\Delta > 0$. A possible realization of spin chains described by  $H_{\rm XXZ}$ is, for instance, a bulk structure of KCuF$_3$ or Sr$_2$CuO$_3$, where the exchange among different chains in the crystal is much weaker than the intra-chain exchange \cite{tennant1995,solog2001,hess2001}. It is well-known that the
Hamiltonian $H_{\rm XXZ}$ can be mapped
onto a LL of spinless fermions \cite{haldane1980,affleck1989,fradkin1991}
\begin{equation} \label{hll}
H_{\rm LL} = \frac{\hbar v}{2} \int dx \left[ g (\Pi(x))^2 + \frac{1}{g} \left(
    \partial_x \varphi(x) \right)^2 \right] ,
\end{equation}
where we have ignored Umklapp scattering \cite{foot1}
and made the identifications
$v=v_B/g$, $v_B = Ja \sin(k_B a)/\hbar$, and $g=(1+4\Delta/\pi)^{-1/2}$ ($a$
is the lattice constant). In Eq.~(\ref{hll}), $\varphi(x)$ is the standard
Bose field operator in bosonization associated with spinon excitations here,
$\Pi(x)$ its conjugate momentum density,
$v$ the spinon velocity, $v_B$ the bare spinon velocity (at $\Delta = 0$), $k_B$ the bare spinon wave vector, and $g$
the interaction parameter ($g=1$ corresponding to a non-interacting system, i.e. $\Delta=0$, and, in general for a $H_{\rm XXZ}$ spin chain, $1/2\leq g\leq 1$) \cite{gogolin1999,giamarchi2004}.
%It is worth noticing that the Hamiltonian $H_{\rm LL}$ can be written as
%\begin{equation} \label{hlc}
%$H_{\rm LL} = \int dx \left[ \frac{(\rho_m(x))^2}{2{\cal C}} + \frac{1}{2}{\cal L} (j_m(x))^2 \right],
%$
%\end{equation}
%where $\rho_m(x)=g_e\mu_B\partial\Phi/\sqrt{2\pi}$ is the local spinon density, $j_m(x)=g_e\mu_B \Pi(x)/
%\sqrt{2\pi}$ the spinon current density, ${\cal C}$ and ${\cal L}$ can respectively
%be then interpreted as the intrinsic spinon capacitance and inductance per unit length \cite{pham}.

In order to be able to properly describe the effect of reservoirs, we modify
the Hamiltonian $H_{\rm LL}$ in the spirit of the inhomogeneous
LL model \cite{maslov1995,ponomarenko1995,safi1995} described by a Hamiltonian $H_{\rm ILL}$, where we assign a spatial dependence
to $v$ and $g$ such that $v(x)=v_l$ and $g(x)=g_l$ being the spinon velocity and the interaction parameter in the reservoirs (for $|x|>L/2$), respectively,
and $v(x)=v_w$ and $g(x)=g_w$ being the corresponding quantities in the spin chain region (for $|x|<L/2$). Within this model, non-equilibrium transport phenomena such as the non-linear
$I-V$ characteristics and the current noise in the presence of impurities
have been analyzed extensively \cite{peca2003,dolcini2003,trauzettel2004,dolcini2005,recher2006}. In this letter,
we are interested in a different situation, namely the ac magnetization transport in the linear response regime which should be seen complementary to the electric ac response analyzed in Ref.~\cite{cuni1998,safi1999a}.

The Hamiltonian $H_B(t)$ describes a spatially varying and
time-dependent magnetic field $\delta B(x) \cos(\omega t) \mathbf{e}_z$ with
$\delta B(x) = -\Delta B/2$ ($\Delta B/2$) for $x<-L/2$ ($x>L/2$). For
$|x|<L/2$, $\delta B(x)$ interpolates smoothly between the values $\pm \Delta
B/2$ in the reservoirs \cite{foot2}. The dc ($\omega=0$) magnetization transport of
such a system has been analyzed in Ref.~\cite{meier2003} and a spin conductance
$G_s=g_l (g_e \mu_B)^2/h$ has been predicted.

The magnetization current in linear response to an oscillating magnetic field can be evaluated using the following expression
\begin{equation} \label{im_linear}
I_m(x,t) = \int_{-\infty}^\infty d \tau \int_{-\infty}^\infty dy \sigma_0(x,y,\tau) \partial_y \delta B(y,t-\tau)
\end{equation}
with
\begin{equation}
\sigma_0(x,y,\tau) = 2i \frac{(g_e \mu_B)^2}{h} \Theta(\tau) \partial_\tau \Bigl\langle [ \varphi(x,\tau), \varphi(y,0) ] \Bigr\rangle
\end{equation}
and the expectation value is taken with respect to $H_{\rm ILL}$. For $x>L/2$ and $|y| \leq L/2$, the spin conductivity is given by
\begin{eqnarray} \label{sigmas}
&& \sigma_0(x,y,\tau) = g_l \frac{(g\mu_B)^2}{h} (1-\gamma) \Theta(\tau) \sum_{p=0}^\infty \sum_{\alpha=\pm} \times \\
&& \Bigl\{ \gamma^{2p} \delta \Bigl( \tau + \alpha \Bigl( \frac{x-L/2}{v_l} + \frac{L/2-y}{v_w} + \frac{2pL}{v_w} \Bigr) \Bigr) + \nonumber \\
&& \gamma^{2p+1} \delta \Bigl( \tau + \alpha \Bigl( \frac{x-L/2}{v_l} + \frac{L/2+y}{v_w} + \frac{(2 p+1)L}{v_w} \Bigr) \Bigr) \Bigr\} \nonumber ,
\end{eqnarray}
where $\gamma = (g_l-g_w)/(g_l+g_w)$
is the reflection coefficient of spinon excitations at a sharp boundary with different interaction coefficients $g_l$ and $g_w$ \cite{safi1995} and $\Theta(\tau)$ the Heaviside function.
%Consequently, the magnetization current at a point $x$ outside the wire is a superposition of different sine waves each of which %weighted by $\gamma^m$, where $m$ corresponds to the number of reflections of the incoming wave at the interfaces between wire and %leads.
The resulting spin current under continuous wave radiation reads
\begin{eqnarray} \label{imfinal}
&& I^{({\rm cw})}_m(x,t) = 2(1-\gamma)\frac{g_l v_w}{\omega} \frac{(g_e \mu_B)^2}{h}
\frac{\Delta B}{L} \sum_{p=0}^\infty \gamma^{2p} \times \nonumber \\
&& \sin \Bigl( \frac{\omega L}{2 v_w} \Bigr) \Bigl\{ \cos \Bigl[ \omega
  \left(t-\frac{(2p+1/2)L}{v_w} + \frac{L-2x}{2v_l} \right) \Bigr] \nonumber \\
&&+ \gamma \cos \Bigl[ \omega
  \left(t-\frac{(2p+3/2)L}{v_w} + \frac{L-2x}{2v_l} \right) \Bigr] \Bigr\} .
\end{eqnarray}
We clearly observe an interaction-dependence of the magnetization current in Eq.~(\ref{imfinal}) through $g_l$ and $\gamma$. The presence of higher harmonics due to higher order terms in $\gamma^m$ would be a strong experimental evidence for the spatial inhomogeneity of spin-spin coupling in realizations of XXZ spin chains. The physics behind the result in Eq.~(\ref{imfinal}) is the following one: the system is driven with a continuous wave due to the ac magnetization source; therefore spinon excitations constantly enter and leave the spin chain from and to the reservoirs. Whenever, they experience a boundary in the exchange interaction, they are partly transmitted and partly reflected with a reflection coefficient $\gamma$. The resulting expression (\ref{imfinal}) is the superposition of all possible contributions to the spin current after infinitely many reflection processes.

As a natural consequence, one may wonder whether an initial magnetization signal is actually transmitted through the spin chain. This depends crucially on the value of $\gamma$. To answer this question, we look at the magnetization current in linear response to a unit pulse described by $\partial_y \delta B(y,t) = \delta B_p \delta \tau_p \delta(t-t_0) \delta(y-y_0)$ with $y_0 \in [-L/2,L/2]$ (where $\delta B_p$ corresponds to the height and $\delta \tau_p$ to the duration of the pulse). If we plug this expression into Eq.~(\ref{im_linear}), we obtain for the spin current
\begin{eqnarray} \label{imfinal_up}
&& I^{({\rm pul})}_m(x,t) = g_l \delta B_p \delta \tau_p \frac{(g_e \mu_B)^2}{h} (1-\gamma ) \Theta(t-t_0) \sum_{p=0}^\infty \times \nonumber \\
&& \sum_{\alpha=\pm} \Bigl\{ \gamma^{2p} \delta \Bigl( \tau + \alpha \Bigl[ \frac{x-L/2}{v_l} + \frac{L/2-y_0}{v_w} + \frac{2pL}{v_w} \Bigr] \Bigr) + \nonumber \\
&& \gamma^{2p+1} \delta \Bigl( \tau + \alpha \Bigl[ \frac{x-L/2}{v_l} + \frac{L/2+y_0}{v_w} + \frac{(2 p+1)L}{v_w} \Bigr] \Bigr) \Bigr\} . \nonumber \\
\end{eqnarray}
The derivation of $I^{({\rm pul})}_m(x,t)$ demonstrates that the initially sharp $\delta$-pulse is decomposed into a sum of infinitely many $\delta$-pulses. Importantly, the amplitude of these pulses decreases by a factor $\gamma$ in a stepwise fashion once in each time interval $L/v_w$ corresponding to the transit time in the wire. So, to answer the question how much signal has been transmitted we have to fix $x$, $y_0$, $t_0$, and $\gamma$ in Eq.~(\ref{imfinal_up}) and sum up all the prefactors of the $\delta$-functions that can be non-zero in a given time interval between $t_0$ and $t$. This analysis implies that all the dissipation happens in the leads and intrinsic relaxation is absent which is a direct consequence of the fact that the LL Hamiltonian describes a free boson. Once we introduce impurities the situation is different and intrinsic dissipation matters which we will briefly address below.

\begin{figure}

\vspace{0.3cm}

\begin{center}
\leavevmode \includegraphics[width=7.5cm]{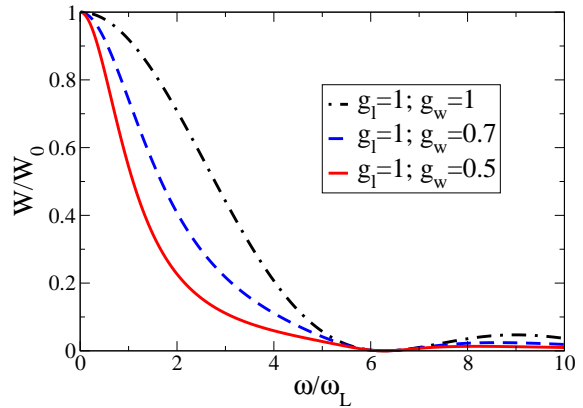}
\caption{(color online) The absorbed power is shown in units of $W_0 \equiv (g_e
  \mu_B \Delta B)^2/2h$ as a function of the ac frequency $\omega$ in units of
the ballistic frequency $\omega_L=v_w/L$. It is clearly visible that stronger
repulsive interactions inside the wire decrease the absorbed power in the system.}
\label{fig_wofom}
\end{center}
\end{figure}

We now turn to the discussion of the power absorption under continuous wave radiation. It turns out that the absorbed power is an ideal physical quantity to measure the reflection coefficient $\gamma$.
We derive the absorbed power of the 1D spin chain using Fermi's golden rule and linear response theory where particular care has to be taken of the spatial inhomogeneity of systems. The resulting expression is
\begin{equation} \label{W1}
W(\omega) = \frac{1}{2} \left\{ \int_{-L/2}^{L/2} dx
\int_{-L/2}^{L/2} dy {\rm Re} \, \sigma_0(x,y,\omega) \right\} \left|
\frac{\Delta B}{L} \right|^2 ,
\end{equation}
where
\begin{eqnarray} \label{resig}
{\rm Re} \, \sigma_0(x,y,\omega) &=& g_w \frac{(g_e \mu_B)^2}{h} \Bigl\{ \cos({\tilde{\omega}} ({\tilde{x}}-{\tilde{y}})) \\
&+& \frac{2 \gamma (1-\gamma^2) \cos ({\tilde{\omega}}) \cos({\tilde{\omega}} ({\tilde{x}}+{\tilde{y}}))}{1+\gamma^4-2\gamma^2
\cos(2{\tilde{\omega}})} \nonumber \\
&+& \frac{2 \gamma^2 \cos({\tilde{\omega}} ({\tilde{x}}-{\tilde{y}})) (\cos(2 {\tilde{\omega}}) - \gamma^2)}{1+\gamma^4-2\gamma^2
\cos(2{\tilde{\omega}})} \Bigr\} , \nonumber
\end{eqnarray}
and we have introduced dimensionless variables ${\tilde{x}} = x/L$, ${\tilde{y}} = y/L$, and ${\tilde{\omega}}=\omega/\omega_L$ with $\omega_L=v_w/L$.
%(Note that Eq.~(\ref{resig}) differs from the real part of the Fourier transform of Eq.~(\ref{sigmas}). This is due to the %different parameter regimes of $x$ and $y$ in the expressions of the spin conductivity that are relevant for the calculation of %$I_m(x,t)$ in comparison to $W(\omega)$.)
It is straightforward to do the two remaining integrals in Eq.~(\ref{W1}) and the final result reads
\begin{eqnarray} \label{W_result}
W(\omega) &=& g_w \frac{(g_e \mu_B \Delta B)^2}{2h} \left( \frac{\sin
({\tilde{\omega}}/2)}{{\tilde{\omega}}/2} \right)^2 \\
&\times& \frac{1-\gamma^4 + 2 \gamma (1-\gamma^2)
\cos({\tilde{\omega}})}{1+\gamma^4-2\gamma^2
\cos(2{\tilde{\omega}})} . \nonumber
\end{eqnarray}
This is the main result of our work. It demonstrates that a measurement of the absorbed power due to ac response of the quantum spin chain is a simple and feasible way to measure interaction dependent coefficients such as $g_w$ and $\gamma$. In Eq.~(\ref{W_result}), these coefficients just appear as pre-factors and not in a complicated power-law fashion as it is usually the case in observable quantities of systems described by LL physics. In Fig.~\ref{fig_wofom}, we show the interaction dependence of the absorbed power $W(\omega)$. This demonstrate that stronger repulsive interactions inside the wire with respect to the leads suppress the dissipative power.

If we compare Eqs.~(\ref{imfinal}) and (\ref{W_result}) we observe an interesting finite-size effect, namely that $W(\omega)$ vanishes as $(\sin ({\tilde{\omega}}/2))^2$ close to $\tilde{\omega} \approx 2 \pi$ whereas the leading contribution to $I_m^{({\rm cw})} (x,t)$ vanishes only as $\sin ({\tilde{\omega}}/2)$ close to that driving frequency. Thus, the power absorption is more strongly suppressed than the magnetization current at frequencies close to $2 \pi \omega_L$. This feature can be used in future devices to transfer data at special frequencies with low power dissipation.
%(Note that the only interaction-dependence of the latter effect is the dependence of $\omega_L$ on $v=v_B/g$.)
%
%%%%%%%%%%%%%%%%%%%%%%%%%%%%%%%%%%%%%%%
%%%%%%%%%%%%%%%%%%%%%%%%%%%%%%%%%%%%%%%
%%%%%      FIGURE    W of omega  %%%%%%
%%%%%%%%%%%%%%%%%%%%%%%%%%%%%%%%%%%%%%%
%%%%%%%%%%%%%%%%%%%%%%%%%%%%%%%%%%%%%%%
%
In the limit $\omega \rightarrow 0$, we obtain $W(0) = g_l (g_e \mu_B \Delta B)^2/2h$ corresponding to Joule heating where $W(0) = I_m^2(\omega=0)/(2G_s)$.

We now address the robustness of our main result (\ref{W_result}) against impurity scattering. An impurity can be modeled as an altered link in the $H_{\rm XXZ}$ chain, i.e. a local change in $J$ on a nearest-neighbor link \cite{kane1992,eggert1992}. Within bosonization, such a scatterer at position $x_0$ in the system can be written as $H_I = \lambda \cos [ \sqrt{4 \pi} \varphi(x_0,t) + 2 k_B x_0 ]$. If one of the two energy scales $\hbar v_w/L$ or $\hbar \omega$ is larger than $\lambda$ (the bare impurity strength), we can treat $H_I$ perturbatively up to lowest non-trivial order (which is second order in $\lambda$). In the presence of impurity scattering, the spin conductivity that enters into the calculation of Eq.~(\ref{W1}) is subject to a (small) correction $\sigma_I(x,y,\omega)$ which has been derived for the corresponding electric case in Ref.~\cite{dolcini2005}. For finite frequencies, $\sigma_I(x,y,\omega)$ needs to be evaluated numerically. In the zero frequency limit, one finds power-law corrections to the spin conductance \cite{maslov1995b,safi1999b} resulting in power-law corrections to the absorbed power. In any case, as long as either $\hbar v_w/L$ or $\hbar \omega$ are larger than the local change in $J$ of the sample, the effect of impurity scattering is weak.

The system which we considered previously consists of a spin chain smoothly connected to
reservoirs. One may wonder how the previous result gets modified for isolated finite size spin chains, to which a time-dependent oscillating magnetic field is applied along the chain (such that $dB(x,t)/dx= \Delta B\cos(\omega t)$).
For long Heisenberg chains, $H_{\rm XXZ}$ still maps onto a LL of spinless fermions as in Eq. (\ref{hll}) but with open boundary conditions (OBC). Following Ref.~\cite{mattsson}, we can establish that
${\rm Re}\, \sigma_0^{OBC}(x,y,\omega) = 2g_w\frac{(g_e\mu_B)^2}{h}\sin(\omega x/v_w)\sin(\omega y/v_w)$
for $\omega=\omega_n \equiv \pi n v_w/L$ ($n=1,\cdots, (L-a)/a$) and $0$ otherwise.
From this expression, we can infer directly the power needed to spatially shake the spin chain, using Eq.~(\ref{W1}),
\begin{equation} \label{W_OBC}
W_{\rm OBC}(\omega) = g_w \frac{(g_e \mu_B \Delta B)^2}{h} \left( \frac{\sin
({\tilde{\omega}}/2)}{{\tilde{\omega}}/2} \right)^2 \sin^2({\tilde{\omega}}/2),
\end{equation}
for $\tilde \omega= \omega_n/\omega_L$ and $0$ otherwise.
Note that this power cannot be identified as dissipative power because a disconnected LL does not contain a dissipative term. Instead, $W_{\rm OBC}(\omega)$ is the work per unit time needed to shake the system. This is the major difference to the case with leads, i.e. Eq.~(\ref{W_result}), where dissipation happens in the reservoirs.
In the limit $\omega\to 0$, $W_{\rm OBC}\to 0$ due to the absence of reservoirs.

Let us now compare typical values for the absorbed power in electric systems
versus magnetic systems. We set $g_l=g_w=1$ for simplicity but keep in mind how finite interactions change the power absorption according to Eq.~(\ref{W_result}).
The absorbed electric power in the dc limit is given by $W_{\rm el}=(e \Delta V)^2/h$.
For a typical electric bias of $\Delta V=1$mV, we obtain $W_{\rm el}\approx 3.87 \cdot 10^{-11}$Js$^{-1}$ whereas the absorbed magnetic power for a typical magnetic bias of $\Delta B=0.1$T is
$W \approx 2.59 \cdot 10^{-15}$Js$^{-1}$ (assuming $g_e=2$) which is four orders of magnitude smaller.
The rule of the thumb is $W_{\rm el}(\Delta V=0.1 {\rm mV}) \sim W (\Delta B=1 {\rm T})$. Thus, we expect substantial advantages of magnetic systems versus electric systems as far as power consumption is concerned.

In summary, we have analyzed the magnetization current and the power absorption of quantum spin chains coupled to magnetization reservoirs with a time-dependent magnetic field applied to the reservoirs. Both physical quantities depend crucially on the difference of the exchange interactions within the wire as compared to the magnetization leads. In fact, we envision to use this dependence as a way to control power dissipation in non-itinerant quantum systems in which magnetization transport occurs via spinons. Finally, we have briefly described the case of a finite size chain and the influence of impurity scattering on spin current.

We acknowledge financial support by the Swiss NSF, NCCR Nanoscience, and JST ICORP.


\begin{thebibliography}{27}
\bibitem{foot1}
The Umklapp scattering term $H_U = \lambda_U \int dx \cos (\sqrt{16 \pi}
\varphi(x))$, where $\lambda_U = -\hbar v_B/(\pi^2 a^2)$, that we have ignored
in Eq.~(\ref{hll}), is an irrelevant term
(in a renormalization group sense) for $1/2 < g \leq 1$ which corresponds to $\Delta<1$.
\bibitem{foot2}
We assume that $1/\omega$ is a long time-scale as compared to typical relaxation times in the reservoirs such that a local thermal equilibrium is established in both reservoirs. Furthermore, we request that $\omega \ll \omega_c = v_B/a$, where $\omega_c$ is the high-energy cutoff frequency of the LL model. Additionally, we assume that the wire is isolated and the spinon number is conserved.
\end{thebibliography}
\end{document}